\documentclass[11pt]{article}
\usepackage{graphics} 
\usepackage{float} 
\usepackage{amsmath} 
\usepackage{amssymb}
\begin{document}
\title{ 
Disks on a Sphere and two-dimensional Glasses}
\author{Werner Krauth
\footnote{krauth@lps.ens.fr,
http://www.lps.ens.fr/$\tilde{\;}$krauth} \\
CNRS-Laboratoire de Physique Statistique \\
Ecole Normale Sup{\'{e}}rieure,\\
24, rue Lhomond, 75231 Paris Cedex 05, France}
\maketitle
\begin{abstract}
Talk given at the conference on `Inhomogeneous Random Systems' at 
the University of Cergy-Pontoise, France, (23 January 2001).\\
\\
I describe the classic circle-packing problem on a sphere, and the
analytic and numerical approaches that have been used to study it.
I then present a very simple Markov-chain Monte Carlo algorithm, which
succeeds in finding the best solutions known today. The behavior of the
algorithm is put into the context of the statistical physics of glasses.\\
\\
keywords: Monte Carlo Methods;  Packing of Circles;  Glass Transition  \\
\\
AMS numbers: 00A72; 52C15; 82D30 \\
\end{abstract}

What is the minimal radius $R_N$ of a sphere allowing $N$ non-overlapping
circles of radius $r=\frac{1}{2}$ to be drawn on its surface?  In a classic
paper, Sch\"{u}tte and van der Waerden \cite{Schuette1} reduced this
problem to a study of planar graphs: As the sphere has minimal radius,
many circles are blocked, i.e. they cannot move. These circles make up the
$M \le N$ vertices of the graph. If two circles are in (blocking) contact,
they are connected by an edge.
\begin{figure}[htbp] 
\begin{center} 
\includegraphics{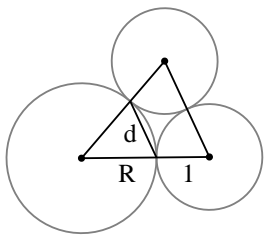} 
\caption{ For $N=13$ spheres, $ 1 < \frac{R}{d} = \frac{R+1}{2}$ as
proven by Sch\"{u}tte and van der Waerden. 
This means that $13$ identical spheres cannot be placed on top of 
another one, of same size.}
\label{f:13spheres} 
\end{center} 
\end{figure}

\begin{figure}[htbp]
\begin{center}
\includegraphics{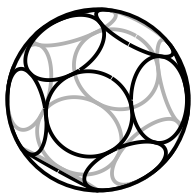}
\end{center}
\caption{ Best known arrangement of $13$ circles of radius $\frac{1}{2}$
(on the surface of a sphere of radius $1.04557...$). As a consequence, 
$13$ spheres of radius $1$ can be placed on the surface of a central 
sphere of radius $r \ge 1.0911\ldots$.}
\label{f:13projection} 
\end{figure}

Sch\"{u}tte and van der Waerden considered
all graphs with $M \le 12$ vertices. Some of these graphs can 
correspond to a blocked configuration of circles on a sphere of radius
$\tilde{R}$, with $R=\min_{\text{graphs}}\tilde{R}$. 
They were able to determine 
$R_N$ for $N \le 12$. 
For this latter case $N=12$, {\em e.g.}, the centers of the circles are located
at the vertices of an icosahedron of side length $1$, and lie on a sphere
of radius 
\begin{equation}
 R_{12} = \frac{\sqrt{10 + 2 \sqrt{5}}}{4} = 0.951\ldots
\label{e:icosahedron}
\end{equation}
Sch\"{u}tte and van der Waerden \cite{Schuette2} were later able to prove
that $R > 1$ for $N > 12$. This result ({\em cf} figure~\ref{f:13spheres})
implies that $13$ spheres of radius $1$ cannot be placed on top of another
sphere of the same size, and solves a packing problem discussed at least
since the time of Newton.  The best currently known packing of spheres
for $N=13$ concerns spheres of radius $1$ on top of another sphere of
radius $1.0911\ldots$.  The corresponding circle arrangement is shown
in figure~\ref{f:13projection}, and is almost certainly optimal. Today,
it should be easy to enumerate planar graphs on a computer, and to prove
this conjecture.

The problem of packing disks on the surface of a sphere has appeared
repeatedly in the literature, as there are important applications in
mathematics  and the natural sciences.  Hundreds of papers have been
written on this subject (for a partial list of references {\em cf}
{\em e.g.} \cite{Conway}), most of them improving upper bounds on $R_N$
in the range of $ 13 < N \simeq 100$. Only for $N=24$ has a new optimal
solution been proven \cite{Robinson}.

The problem of $N$
non-overlapping disks on a minimal sphere can be reformulated  as
a minimization problem of the potential energy of $N$ particles
$i=1,\ldots,N$ on the surface of the sphere of radius $1$ with a two-body
interaction \cite{Clare,Kottwitz}
\begin{equation}
V = \sum_{i=1}^{N-1} \sum_{j=i+1}^{N} \left(\frac{c}{r_{ij}}\right)^p
\label{e:}
\end{equation}
where $r_{ij}$ is the Euclidean  distance between  circles $i$ and $j$.
Therefore, customary minimization  programs (such as the Newton-Raphson
method) can be applied to find local - and
hopefully global - minima for ever increasing values of $p$. In our first
units (circles of radius $\frac{1}{2}$ on a sphere of radius $R$),  the minimal
radius $R_N$ is given by
\begin{equation}
R_N = \frac{1}{\min_{ij} r_{ij}}
\label{e:potential}
\end{equation}
in the limit $p \rightarrow \infty$. 

Current records were set by Kottwitz \cite{Kottwitz}, who used values
of $p$ in the astonishing range of $ p=80$ to $p= 1\,310\,720$.  Running
such a sophisticated Newton-Raphson program is  a delicate task as one
has to avoid getting trapped in local minima of the potential energy.
Nevertheless, the strategy proved to be successful, as a large number of
new values for $R_N$ were found, and as these values have so far stood
the test of time.

Recently, I have reconsidered the non-overlapping circle problem from
the view point of Markov-chain Monte Carlo algorithms.  Instead of
searching for the minimum of a regularized energy, I move circles during
a large number of time steps $t=1, \ldots,T_{\max}$. At time $t$, a
single, randomly chosen  circle $i$ is moved a tiny bit into a random
direction. Specifically, let us denote the position of $i$ at time $t$
by $\vec{r}_i$, and let the displacement vector  be $\vec{\delta r} =
(\delta x, \delta y, \delta z)$ with Gaussian random numbers $ \delta x$
{\em etc.} that have zero mean and variance $\sigma^2 \ll 1$. Then the
following move is considered \cite{footgauss}
\begin{equation}
\vec{r}_i \rightarrow  
\vec{r}'_i = 
\frac{\vec{r}_i + \vec{\delta r} }{|\vec{r}_i + \vec{\delta r}|} R
\label{e:}
\end{equation}
If the move can be performed without violating the constraints, it is
accepted, and rejected otherwise. This procedure implements a  standard
{\em local} Metropolis algorithm for a constant stationary probability
distribution: Any non-overlapping configuration of circles
should be visited with the same probability.

In addition to performing the Metropolis algorithm at fixed $R$, I also
implemented the ``simulated annealing'' procedure \cite{Kirkpatrick}:
Every so often,  the radius of the sphere is reduced by a very small
amount, if this is possible without introducing overlaps between particles
\cite{request}.

I have found that this simple  algorithm reproduces the best
results known up to date ({\em cf} \cite{Kottwitz}) for $N=5 \ldots 90$,
typically within a few minutes of computer time \cite{footprec}.

This observation may be of practical interest, as the Monte Carlo
algorithm is much simpler than the previous approaches. However, it also
underlines the quality of the bounds obtained by Kottwitz, which I have
been able to reach, but not to improve on.

In physical terms, the graph giving $R_N$ is the ground state of the
system (which does not have to be unique: for $N=15$, {\em e.g.},
two best solutions are known \cite{Kottwitz}).  Following Stillinger
and Weber \cite{Stillinger}, all the other graphs that can be put on
the surface of a sphere (with larger radii $\tilde{R}> R_N$) may be called
`inherent structures'. These are the configurations that the simulated
annealing algorithm may get trapped in. If the Monte Carlo algorithm
is trapped in an inherent structure at radius $\tilde{R}$, it could by
continuity also get trapped by the same inherent structure at a slightly
larger radius $R' \stackrel{>}{\sim} \tilde{R}$. This means that, 
in the strict sense,
the Monte Carlo algorithm is non-ergodic at least for all $R'$ satisfying
\begin{equation}
        \max_{\text{graphs}}\tilde{R} > R' > R_N.
\label{e:}
\end{equation}
In this and similar systems, very little is known rigorously about the
number of inherent structures as a function of radius  (their density
of states), and enumeration methods at small $N$ should be very helpful.
For physical applications, the thermodynamic limit
$N \rightarrow \infty$ is of prime interest. In this limit, $(R' -
\tilde{R})/\tilde{R}$ certainly goes to zero. This means that strict
ergodicity (on timescales that diverge with $N$) is reinstalled in the
thermodynamic limit. Precise understanding of this complicated thermodynamic
limit is lacking. 

If, in contrast, the simulated annealing algorithm got trapped in a
non-optimal state (with $R$ clearly larger than $R_N$) with probability
close to one, we would say that the `physical' system  of disks on the
surface of a sphere had a glassy phase. Such a phase seems to be absent
for monodisperse circles. Our findings agree very well with what has
been found in monodisperse hard disk and hard sphere systems on two-
and three-dimensional tori \cite{Santen,Torquato}.

In contrast, there is strong numerical evidence that {\em poly}disperse
hard disks can be glassy: in a  system of hard disks of varying radii
on a torus \cite{Santen}, a comparable local Monte Carlo algorithm
{\em always} gets trapped at finite density. Again, it is fundamentally
unclear how this observation carries over into the thermodynamic limit.
Empirically, diffusion times of particles seem to grow without bounds
at a well-defined, finite density.

In these two-dimensional systems, it has been found \cite{Santen} that
other Monte Carlo dynamics do not suffer the same slow-down as the local
algorithm. These methods also provide strong indications that the glass
transition in these two-dimensional models is a purely dynamic phenomenon,
which does not manifest itself in equilibrium averages over the stationary
probability distribution.

Acknowledgements: 
I would like to thank L.~Santen and S. Tanase-Nicola for helpful discussions.


\begin{thebibliography}{10}
\bibitem{footgauss} It is also possible to orthogonalize $\vec{\delta r}$ with respect
to $\vec{r_i}$.
\bibitem{request}
The resulting algorithm has a few dozen lines and can be
obtained on request.
\end{thebibliography}
\end{document}